# Emergent superconductivity at 16.3 K in an altermagnetic candidate Na$_{2-x}$V$_2$Se$_2$O with broken inversion symmetry

Yingrui Sun[1,2,§], Zhaocheng Yin[1,2,§], Tengfeng Zhang[1,2,§], Lihan Wang[1,2], Binbin Ruan[3], Yu Huang[1,2], Junbao He[4], Wenliang Zhu[5], Maoyun Ma[1,2], Jianli Bai[1,2], Jingwen Cheng[1,2], Qingxin Dong[1,2], Cundong Li[1,2], Pinyu Liu[1,2], Qiaoyu Liu[1,2], Chengdong Zhang[1,2] and Genfu Chen[1,2,*]

[1]Institute of Physics and Beijing National Laboratory for Condensed Matter Physics, Chinese Academy of Sciences, Beijing, 100190, China

[2]School of Physical Sciences, University of Chinese Academy of Sciences, Beijing, 100049, China

[3]College of Physics and Center of Quantum Materials and Devices, Chongqing University, Chongqing, 401331, China

[4]College of Physics and Electronic Engineering, Nanyang Normal University, Nanyang, 473061, China

[5]School of Physics and Information Technology, Shaanxi Normal University, Xi'an 710119, China

*Corresponding author. E-mail: gfchen@iphy.ac.cn

**ABSTRACT**

**Altermagnets (AMs), characterized by zero net magnetization and momentum-dependent spin splitting, are anticipated to hold significant potential for generating multiple exotic and uncommon superconducting states. However, superconductivity has not yet been realized in AMs to date. Recently, two-dimensional (2D) V$_2$*Ch*$_2$O (*Ch* = Se, Te) monolayers, as well as *A*V$_2$*Ch*$_2$O (*A* = K, Rb, Cs) crystals containing [V$_2$*Ch*$_2$O]$^{δ-}$ building layers, have been predicted and/or demonstrated to be promising altermagnetic materials. Our**

**preliminary attempts to explore superconductivity in these materials by applying pressure or chemical doping were unsuccessful. Here we report the discovery of superconductivity at a relatively high transition temperature of ∼ 16.3 K in a newly synthesized layered compound, $Na_{2-x}V_2Se_2O$, a variant of $AV_2Ch_2O$. In this structure, the $[V_2Ch_2O]^{\delta-}$ layers are interspersed with double layers of $Na^+$ instead of a single layer of $A^+$, with sodium sites being only half-filled. This new family of layered vanadium oxychalcogenides, lacking inversion symmetry, represents an intriguing platform for exploring altermagnetic superconductors, and holds the potential to reveal novel phenomena, such as topological states, van Hove singularities, and finite-momentum superconductivity. Furthermore, this material acts as a "bridge" between the cuprate/nickelate and iron-pnictide high temperature superconductors, providing new hope and opportunity to expand the category of layered superconductors with higher critical temperatures ($T_C$) and enhancing our understanding of the underlying mechanisms in these systems**.

## INTRODUCTION

Despite extensive searches over the past century, only a few families of bulk high-temperature superconducting materials at ambient pressure are available, and the nature of the pairing mechanism remains unclear [1,2]. Nevertheless, it has been well-established that the existence of two-dimensional (2D) conducting layers, such as $[CuO_2]$ or $[Fe_2X_2]$ ($X$ = Pnictogens or chalcogens), modulated by the blocking layers that serve as charge reservoirs, is critical for achieving high-$T_C$ superconductivity. In these systems, the superconducting ground state emerges from a parent antiferromagnetic state through carrier doping or pressure application. Therefore, a feasible strategy to explore new superconductors with potentially high-$T_C$ is to design and synthesize novel 2D layered materials that incorporate transition metals and host antiferromagnetism.

$AV_2Ch_2O$, a novel 2D vanadium oxyselenide, possesses an anti-perovskite structure

containing an inverse $V_2O$ square lattice with a low mixed oxidation state ($V^{2.5+}$, $3d^{2.5}$). In this structure, the $V_2O$ plane is anti-structural to the $LaNiO_2$-type $CuO_2$ net found in cuprate superconductors [3,4]. Additionally, chalcogen anions located above and below the center of the V-O square form a structural motif similar to that found in iron-based superconductors. Consequently, the vanadium ion has an octahedral $VO_2Ch_4$ coordination, and the $V_2Ch_2O$ octahedra share edges to form a two-dimensional network. This close structural relationship to both cuprates and iron pnitrides enhances the attractiveness of this system, leading to the emergence of peculiar physical properties in $AV_2Ch_2O$. For instance, it has been observed that this system exhibits a spin-density-wave (SDW)-like transition at low temperature and undergoes an antiferromagnetic phase transition at higher temperatures [3,4]. These characteristics position it as a promising candidate for achieving high-$T_C$ superconductors. Very recently, further systematic characterization and analysis of the magnetic and electronic structures have challenged the simple antiferromagnetic picture, instead suggesting the existence of an altermagnetic order with $d$-wave spin splitting in the electronic bands − a feature reminiscent of a $d$-wave superconducting order parameter [3-6].

This structural motif was first reported in titanium-based pnictide oxides three decades ago [7], shortly after the discovery of cuprates. Subsequently, layers with similar arrangements of $d$-metal, chalcogen, and oxygen have been synthesized one after another, such as $La_2O_2Fe_2Se_2O$, $Na_2Fe_2Se_2O$, $La_2O_2Co_2Se_2O$ [8-10]. These materials gained renewed interest following the discovery of iron-based superconductors. In contrast to bad-metal properties of the parent iron-pnictides, the Fe- or Co- based oxychalcogenides exhibit Mott-insulating behavior with a small band gap [11-13]. Efforts involving pressure application or chemical doping have been carried out on these materials with the aim of discovering new superconductors. Unfortunately, to date, superconductivity has only been realized in $(Ba,Na)Ti_2(Bi,Sb)_2O$ (containing $Ti^{3+}$ ions with $3d^1$ configuration) in proximity to charge-density-wave (CDW) instability, and the transition temperature is relatively low ($T_C < 6$ K) [14,15]. The underlying mechanism of the superconductivity in this material is still an active topic of debate, though charge fluctuations may play an important role in the pairing mechanism [16-18]. Finding parent structures with $d^1$ square lattices was once considered a feasible strategy for obtaining new high-$T_C$ superconductors, as the $d^1$ state is regarded as an electron-hole symmetric state of $3d^9$ state in cuprates. Whether a variant of this material will exhibit a higher $T_C$ is yet to be explored. This also raises the question of whether superconductivity with a significantly higher $T_C$ can occur in such layered anti-perovskites that do not contain titanium.

$A$V$_2$$Ch$$_2$O is unique in that it serves as a metallic room-temperature altermagnet, accompanying an SDW-like transition at low temperatures [4-6]. Remarkably, small effective mass, high mobility and nonzero Berry phases detected in KV$_2$Se$_2$O provide direct evidences for the existence of topologically nontrivial electronic structures in this family [3], consistent with the theoretical predictions of Weyl semimetal states and nodal lines in altermagnets [19,20]. Moreover, $A$V$_2$$Ch$$_2$O contains V$^{3+}$/V$^{2+}$ ($3d^2$/$3d^3$), which corresponds to the electron analogue of the formally Ni$^{2+}$/Ni$^{3+}$ ($d^8$/$d^7$) layered nickelate, i.e., La$_3$Ni$_2$O$_7$, a newly discovered superconductor containing 2 or 3 holes in the Ni $d$-shell [21,22]. All these imply that $A$V$_2$$Ch$$_2$O may exhibit a higher critical temperature and unconventional superconductivity if spin fluctuations trigger superconductivity and if, like iron pnictides, the SDW/AM states could be suppressed upon chemical doping or pressure application. However, initial attempts using these tuning routes have not yielded any signs of superconductivity. Nevertheless, it might be possible to tune V$_2$$Ch$$_2$O layer into superconducting state by modulating the valence of vanadium to form states closes to V$^{2+}$ ($3d^3$) or V$^{3+}$($3d^2$), and/or by optimizing the interlayer spacing—the distance between neighboring V$_2$$Ch$$_2$O layers—as seen in titanium-based pnictide oxides [14,15]. In those cases, superconductivity occurs only when the two adjacent [Ti$_2$$Pn$$_2$O]$^{2-}$ ($Pn$ = P, As, Sb, Bi) layers are separated by single-layer Ba$^{2+}$ ions. No superconductivity was observed in titanium-based pnictides when the single-layer Ba$^{2+}$ ions were replaced with double layers of Na$^{2+}$, or [Sr$_2$F$_2$]$^{2+}$ [7,23].

On the other hand, despite rapid progress in understanding altermagnetic materials—such as CrSb, $\alpha$-MnTe, and CoNb$_4$Se$_8$ [24-27], experimental realizations of altermagnetic superconductivity remain elusive, and the existence of superconductivity in these materials is still doubtful. Superconductivity is often absent or weak in altermagnetic materials because their superconducting states rely intrinsically on interband pairing of electrons from bands split by spin-orbit coupling (SOC) [28, 29]. Nevertheless, applying an external magnetic field in strong altermagnets, or chemical doping in altermagnetic insulators are the promising routes for the realization of altermagnetic superconductivity [30]. Consequently, an intriguing question arises regarding what type of superconducting symmetry might emerge from the AM-type spin fluctuations. Theoretical proposals indicate that altermagnetic superconducticity could stabilize spin-triplet pairing, topological superconductivity, or even exotic phases such as the Fulde-Ferrell-Larkin-Ovchinnikov (FFLO) state [31-36]. Therefore, finding an optimal altermagnet candidate and successfully inducing superconductivity are crucial steps toward unraveling these mysteries.

In this work, we report the discovery of a new family of layered vanadium oxychalcogenides, $Na_{2-x}V_2Se_2O$, with a non-centrosymmetric structure. Similar to $AV_2Ch_2O$, the structure of $Na_{2-x}V_2Se_2O$ features the alternating stacking of $[V_2Se_2O]^{\delta-}$ layers and $[Na_{2-x}]^{\delta+}$ layers. However, it involves double layers of $Na^+$ instead of a single layer of $A^+$. The enlarged spacing between two neighboring $[V_2Se_2O]^{\delta-}$ layers enhances its two dimensionality and substantially weakens interlayer exchange coupling. $Na_{2-x}V_2Se_2O$ exhibits an SDW-like anomaly at about 85 K, with emergence of superconductivity up to 16.3 K. Notably, the low-temperature resistivity upturn observed in the parent materials closely resembles that reported for under-doped cuprates, iron oxypnictides and nickelates. First-principles calculations emphasize the quasi-two-dimensional nature, as well as the Dirac-like energy dispersions near Fermi level. At present, the superconducting volume fraction of the polycrystalline samples prepared by solid-state reaction or topochemical approaches, as well as as-grown single crystals, is only about 5% (or lower). This indicates that further work is required to pinpoint the optimal doping level. Nevertheless, our discovery not only opens a door to realize rather high-$T_C$ superconductivity in altermagnets, but also provides a unique opportunity to explore the interplay between superconductivity and other exotic states, such as SDW, in the context of altermagnetism. Furthermore, by leveraging structural flexibility, new superconductors with various stacking structures could be designed by modifying the blocking layers.

**RESULTS & DISCUSSION**

**Structure Characterization**

In contrast to the tetragonal structure ($P4/nmm$) of $A$V$_2$$Ch_2$O [3], Na$_{2-x}$V$_2$Se$_2$O crystalizes in the orthorhombic space group $Ammm$ (No.65) at room temperature. As shown in Fig. 1(a), the $a = b$ lattice parameter is insufficient to define tetragonal symmetry in this system. The distinction between the two crystal structures is also reflected in the morphology of the obtained single crystals: $A$V$_2$$Ch_2$O crystals are square and plate-like, whereas Na$_{2-x}$V$_2$Se$_2$O crystals form elongated, lath-like shapes. Interestingly, this distortion closely resembles that observed in the copper- and iron-based systems, where the parent compounds undergo a subtle, cooling‑induced distortion that breaks tetragonal symmetry and drives a transition to an orthorhombic phase [37,38]. Since that structural change effectively relieves magnetic frustration, it is considered to be magnetically driven. Upon doping or applying pressure, the structural transition is suppressed and superconductivity emerges, but the critical relationship between structure and superconductivity remains unresolved. Remarkably, the parent cuprate material La$_2$CuO$_4$ has recently been proposed as an altermagnet candidate at low temperatures, which is considered to arise from the reduced crystal symmetry due to small rotations of CuO$_6$ octahedra, while the high-symmetric phase at high temperature does not host altermagnetism. Moreover, the resulting spin splitting is small and disappears when the structure becomes tetragonal under doping [39].

While the phase structure changed from tetragonal to orthorhombic with the double Na$^+$-layer instead of single K$^+$-layer, no structural distortion transition has been detected in KV$_2$Se$_2$O, even at temperatures around 105 K, where obvious anomalies in magnetic susceptibility and electrical resistivity were observed [3]. The smaller size of Na means a shorter Pauli repulsion radius, which makes Na prefer to stay even closer to Se as compared to K. It is noteworthy that, in the closely structural model reported for $Na_2Fe_2Se_2O$ and $Na_2Ti_2As_2O$ [7,9], similar types of voids are fully occupied, but there is ~ 50% occupancy of the sodium sites in $Na_{2-x}V_2Se_2O$. It's more like the position of K in $KV_2Se_2O$ splitting into two equivalent atomic sites, each with 50% occupancy. The final formula, $NaV_2Se_2O$, is consistent with EDX and ICP analyses on the obtained single crystals, indicating that the reduction of $V^{2.5+}$ is not easy to attain in the $V_2Se_2O$ block, even though excessive sodium was used during the sample preparation process. Why the structure favors vacancies on Na sites rather than full occupancy in $Na_{2-x}V_2Se_2O$ is a question worthy of further exploration. Nevertheless, the lattice and bond parameters collectively indicate that the unit cell is chemically compressed in response to variations in cation stoichiometry when compared to $Na_2Fe_2Se_2O$, even though the radius of the Fe ions is smaller than that of the V ions [9]. Such a high concentration of defects in $Na_{2-x}V_2Se_2O$ could also imply that a superstructure with a long-range order of the sodium vacancies is plausible. Compared to a single K-layer, perhaps a double Na-layer is more beneficial for maintaining the stability of the layered framework structure during a limited Na removal process, making it easier to control the carrier concentration than $KV_2Se_2O$. It is, therefore, reasonable to believe that $Na_{2-x}V_2Se_2O$ holds great potential rather than $KV_2Se_2O$ to explore emergent superconductivity in spin-polarized electronic environments.

**Physical Property Characterization of the Parent Material**

Figures 2(a) and (b) display the temperature dependence of the magnetic susceptibility, $\chi(T)$, for $Na_{2-x}V_2Se_2O$, measured with a magnetic field of $\mu_0H = 1$ T applied parallel to the $c$-axis and the $ab$-plane, respectively. The $\chi(T)$ data reveal a significant magnetic anisotropy between these two crystallographic directions. Along the $c$-axis, the susceptibility decreases with decreasing temperature at high temperatures, followed by a sharp jump at $T_{SDW}$, 85 K; this anomaly is reminiscent of the SDW transition observed in $KV_2Se_2O$. Above $T_{SDW}$, the zero-field-cooled (ZFC) and field-cooled (FC) curves essentially overlap in both directions, indicating the absence of hysteresis and a second-order transition. However, the divergence of the FC and ZFC curves below $T_{SDW}$ suggests that $Na_{2-x}V_2Se_2O$ hosts a complex, non-ergodic magnetic state along the *c*-axis, clearly different from that of $KV_2Se_2O$. In contrast, the susceptibility along the *ab*-plane direction exhibits nearly linear temperature dependence above $T_{SDW}$, accompanied by a weak plateau near $T_{SDW}$, analogous to the SDW transition in $KV_2Se_2O$. The minor difference between the ZFC and FC curves is likely attributed to measurement error.

To further investigate the magnetic properties, isothermal magnetization *M*(*H*) curves were obtained at different temperatures, as shown in Figs. 2(c) and (d). For the *c*-axis, the magnetization shows a small initial increase with rising magnetic field, followed by a linear increase at higher fields. However, no hysteresis loop is observed in the low-field region, indicating that the spontaneous net magnetization at zero field is too small to be distinguished. When the magnetic field is applied within the *ab*-plane, the *M*(*H*) curve exhibits quasi-linear behavior. In contrast to $KV_2Se_2O$, these findings imply that the low-temperature ground state of $Na_{2-x}V_2Se_2O$ is probably a canted antiferromagnetic state. Indeed, in a non-centrosymmetric crystal, the combination of SOC and broken inversion symmetry gives rise to the Dzyaloshinskii-Moriya Interaction (DMI). This interaction favors perpendicular alignment of neighboring spins, driving the formation of non-collinear magnetic structures [40,41].

Figure 3(a) shows the temperature-dependent in-plane resistivity $\rho_{ab}$ (*T*) measured on a single crystal sample. The absolute value of $\rho_{300\ K}$ was found to be ~400 μΩ cm, which is on the same order of magnitude as $KV_2Se_2O$. The resistivity decreases with decreasing temperature down to 100 K, displaying a pronounced hump near 85 K. Below 30 K, the resistivity exhibits a rapid increase as the temperature falls. However, in $KV_2Se_2O$, the electron–phonon interaction retraces well the metallic resistivity behavior at low temperatures [3].

Low-temperature resistivity upturn appears to be a common feature among all high-$T_C$ superconductors and remains a critical unresolved issue. In under-doped cuprates, various mechanisms have been proposed to explain this behavior, such as the 2D weak localization effect due to quantum corrections, scattering off $Cu^{2+}$ Kondo impurities induced by residual apical oxygen, and scattering off magnetic droplets induced by disorder [42-47]. Interestingly, newly discovered nickelate superconductors, like $Nd_{1-x}Sr_xNiO_2$, $La_3Ni_2O_7$, and $La_4Ni_3O_{10}$, also exhibit similar anomalies in their parent compounds or under-doped regimes, which are ascribed to the strong correlation effects, or Kondo-like spin–orbit interactions [48-52].

Moreover, applying magnetic field enhances both the magnitude of the upturn and the temperature of the minimum, $T_{min}$ (the temperature at which the resistivity minimum occurs), which seems to rule out the weak localization mechanism as well as the Kondo effect. Given the fact of partially occupancy in $Na_{2-x}V_2Se_2O$, there might be a link between the unusual insulating-like behavior and disorder; the ultimate origins of anomalous upturn in $Na_{2-x}V_2Se_2O$ are yet unclear.

Figure 3(b) displays the field-dependent magnetoresistance ($MR = [\rho(\mu_0 H) - \rho_0]/\rho_0 \times 100\%$) of $Na_{2-x}V_2Se_2O$ at various temperatures. The MR exhibits a quadratic dependence on the magnetic field at low fields and tends to saturate at high fields. At 2 K and 5 T, the magnetoresistance is approximately 14%. Notably, the data in Fig. 3(a) indicate that the MR reaches 100% at 2 K and 16 T, which is significantly higher than that of ordinary metals. Figure 3(c) shows the field-dependent Hall resistivity $\rho_{xy}(\mu_0 H)$ at selected temperatures measured on single crystal sample. The positive slope of $\rho_{xy}$ suggests that hole-type carriers dominate the transport properties in $Na_{2-x}V_2Se_2O$. One can consider the effect of Na-vacancies on the transport properties of the $[V_2Se_2O]^{\delta-}$ planes, which may lead to hole doping of the $[V_2Se_2O]^{\delta-}$ planes in the 2221 type structure, noting that hole-type carriers also dominate the transport properties in $KV_2Se_2O$ with a 1221 type structure with no K vacancies present. Using the single-band model, we extracted the hole concentrations (n) and mobilities (μ) from linear fits to the Hall resistivity data, as shown in Fig. 3(d). At 10 K, the hole concentration is n ≈ 3×10$^{20}$ cm$^{-3}$ and the mobility is $\mu$ ≈ 70 cm$^2$/(Vs). The relatively low mobility and corresponding high electrical resistance indicate that structural defects in Na$_{2-x}$V$_2$Se$_2$O are the primary limiting factor for carrier transport in the low temperature region.

The introduction of Na-vacancies produces additional holes as the V valence changes to accommodate these vacancies. Indeed, attempts to introduce electron doping in Na$_{2-x}$V$_2$Se$_2$O and KV$_2$Se$_2$O were unsuccessful, indicating that the system exhibits a greater tendency towards hole-type conduction. As depicted in Fig. 3(d), the temperature-dependent carrier concentration exhibits a sharp decline below 100 K, implying the opening of an energy gap related to the SDW-like transition, and leading to the resistivity kink, as observed in KV$_2$Se$_2$O [3].

**Emergence of Superconductivity State**

Next, we focus on the emergence of superconductivity in Na$_{2-x}$V$_2$Se$_2$O. As shown in the lower inset of Fig. 4(a), the temperature-dependent in-plane resistivity $\rho_{ab}(T)$ of a single crystal sample grown with different flux ratios is plotted. Besides a peak near 85 K, a distinct drop appears around 10 K, which we attribute to the onset of the superconductivity. Figure 4(a) clearly shows this resistive drop shifts when a magnetic field is applied, confirming its superconducting origin. The estimated $\mu_0 H_{c2}(T)$ -$T$ phase diagram is shown in the upper inset of Fig. 4(a). Having a large upper critical field, even exceeding the paramagnetic limit, is generally considered to be a strong indicator of unconventional superconductivity, especially spin-triplet superconductivity [53]. Superconductivity is observed in the non-centrosymmetric 2D Na$_{2-x}$V$_2$Se$_2$O, but its upper critical field $\mu_0 H_{c2}(0)$ is only moderate. This, however, does not necessarily mean that the superconducting state is of a conventional *s*-wave form. Noted that Na$_{2-x}$V$_2$Se$_2$O crystallizes in a layered structure, therefore the upper critical field should have a large anisotropy.

Alternatively, the observed reduction of $\mu_0 H_{c2}(0)$ in Na$_{2-x}$V$_2$Se$_2$O，is likely ascribed to the inter-band pairing, which is more comparable to the suppression of conventional spin-singlet pairing under magnetic fields [28,54]. Indeed, a strongly suppressed upper critical field has been observed in LaNiGa$_2$ ($T_C$ = 1.8 K; $\mu_0 H_{c2}(0)$ = 0.11 T), where pairing between electrons with the same spins but on different orbitals give rise to a triplet superconducting state [55]. For Na$_{2-x}$V$_2$Se$_2$O, further investigations are necessary to clarify its origin, such as improving the quality of superconducting single crystals and investigating the anisotropy of the upper critical field.

To further confirm the presence of superconductivity indicated by resistivity measurements, DC magnetization studies as a function of temperature were performed on single crystals from the same batch. As shown in Fig. 4(b), the sample shows a strong diamagnetic transition below 10 K, consistent with the electrical measurements. The emergent coexistence of $T_C$ and SDW observed here appears similar to the phenomenon previously seen in parent $A$Fe$_2$As$_2$ ($A$=Ca, Sr, Ba) single crystals, where the behavior has been attributed to lattice distortion or strain acting as an effective pressure to induce superconductivity [56]. However, in our system, external pressure tends to drive the material into an insulating state [unpublished]. Consequently, the observed superconductivity is more likely to originate from a self-doping effect associated with the Na sites.

Usually, polycrystalline samples offer a clear advantage for carrier introduction: the abundant grain boundaries provide rapid diffusion pathways for dopant species such as sodium or oxygen. In addition, polycrystalline synthesis (e.g., solid‑state sintering) does not demand the exact control of temperature gradients and growth conditions required for single crystal growth. Consequently, we can screen many different chemical stoichiometries much more quickly. Figure 4(c) shows the temperature dependence of the resistivity for a polycrystalline sample. Below $T_C$, the resistivity does not decrease smoothly with further cooling; instead, it increases steeply again. Figure 4(d) shows a typical ZFC and FC curves measured on another polycrystalline sample. The onset critical temperature ($T_C$), defined as the temperature at which the magnetization begins to drop, reaches up to 16.3 K. Using the ZFC susceptibility data, the superconducting shielding fraction is estimated to be approximately 5%. Although all prepared polycrystalline samples exhibit a diamagnetic response with $T_C$ ranging from 8 to 16 K, the superconducting volume fraction varies from sample to sample, a behavior also observed in single crystals (See Supplementary Information). All resistance and magnetization measurements show that superconductivity remains deeply intertwined with the magnetic background.

To verify the intrinsic nature of the superconductivity and the role of Na stoichiometry, we performed topochemical intercalation experiments. First, we removed the element K from the non-superconducting $KV_2Se_2O$ polycrystalline sample to obtain $V_2Se_2O$. $V_2Se_2O$ itself showed no superconductivity (see the left inset of Fig. 4(e)), but after soaking it in Na-Naph/THF solution for 18 hours, two superconducting transitions emerged simultaneously at 4 K and 8.5 K, as shown in Fig. 4(e). However, with prolonged soaking, superconductivity completely disappeared, as shown in the right inset of Fig. 4(e). We speculate that this effect is related to the intercalation concentration of Na.

Furthermore, we found that, the low-temperature upturn disappeared and was replaced by a weak superconducting drop, as shown in Fig. 4(f), when we tried to extract the sodium ions quickly from the layered structure by immersing the single crystal in acetone for a few minutes; however, the hump around 85 K became more pronounced. This further suggests that hole doping in such a system is beneficial to the emergence of superconductivity.

At present, preparing high-quality superconducting samples of $Na_{2-x}V_2Se_2O$ remains a significant challenge. Early discoveries of superconductors like $Cs_3C_{60}$ and $Na_xCoO_2$ exhibited very low superconducting volume fractions [57,58]; notably, $Cs_3C_{60}$ initially showed a fraction of only ∼1%. Likewise, $Cu_xBi_2Se_3$, the first candidate topological superconductor obtained by intercalating Cu between the layers of $Bi_2Se_3$, also failed to show zero resistance in its initial reports [59]. Subsequent refinements in synthesis methods substantially increased these fractions, eventually enabling bulk superconductivity. However, superconductivity in altermagnetic materials faces a more fundamental challenge. Theoretical studies predict that superconductivity in these systems is either absent or suppressed, as their superconducting states depend critically on interband electron pairing [28,29], and possible triplet superconducting states are notoriously sensitive to non-magnetic impurities [60]. Furthermore, $Na_{2-x}V_2Se_2O$, suffers from intrinsic inhomogeneity and pronounced disorder, which further hampers the emergence of a coherent superconducting phase. Moreover, when altermagnetic order is embedded in non-centrosymmetric crystal structures, the absence of inversion symmetry permits the DMI to stabilize non-collinear spin textures. The simultaneous breaking of time-reversal, inversion, and rotational symmetries creates a highly correlated electronic environment that is exceptionally sensitive to carrier doping.

In our $Na_{2-x}V_2Se_2O$ samples, self-doping weakens the magnetic exchange just enough for a Cooper-pair instability to develop, yet it does not eliminate the underlying magnetic order. As a consequence, superconductivity remains extremely fragile: it may emerge only in localized regions where magnetism is weak or where particular structural features (such as boundaries or interfaces) exist, occupying only a very small fraction of the sample volume. For a bulk superconducting condensate to emerge, both the SDW and altermagnetic orders must be essentially suppressed.

**First Principles Calculation**

To gain more insight into the electronic states of $Na_{2-x}V_2Se_2O$, we performed density functional theory (DFT) calculations on $Na_{2-x}V_2Se_2O$ ($x = 1$). The results are shown in Figure 5. There are three bands crossing the Fermi level ($E_F$), consistent with the metallic nature. Taking the SOC effect into account does not bring about significant changes to either the band structure [Figure 5(a)] or the density of states [DOS, Figure 5(b, c)], which is as expected since $Na_{2-x}V_2Se_2O$ consists of relatively light elements. The states near $E_F$ mainly originate from the V-3$d$ orbital, as suggested by the projected DOS. Interestingly, there are several van Hove singularities located at ~ 1.0 eV, 0.65 eV, -0.4 eV, -0.85 eV, etc., indicating that the electron correlation could be significant. These van Hove anomalies are closely related to the flat bands at the corresponding energy levels and are originated from the quasi-two-dimensional nature of Na$_{2-x}$V$_2$Se$_2$O. As revealed in Figure 5 (e–g), there are three sheets of Fermi surfaces in the first Brillouin zone. Apart from tiny hole-type pockets near E$_0$ and electron-type pockets near T/C$_0$, the largest part of the Fermi surfaces is a quasi-two-dimensional sheet with rather complex topology. The low-dimensional feature is also reflected in the band structure. For example, the energy dispersions along Γ–Y, T–Z, and C$_0$–S are very small. While the dispersions on perpendicular directions (Y–T, S–R, Y–E$_0$) are large. It is expected that carriers with smaller effective mass and higher mobilities can be observed along these directions. Remarkably, there are several Dirac points (e.g., near Γ/T/C$_0$) on these high-mobility path just around $E_F$, making possible the occurrence of topological superconductivity. In this regard, a more careful examination of the band structure considering the magnetic order and delicate spectroscopic characterizations [e.g., angle-resolved photoemission spectroscopy (ARPES), scanning tunneling spectroscopy (STS)] are definitely needed to elucidate the question.

We note that the band structure and Fermi surfaces of Na$_{2-x}$V$_2$Se$_2$O ($x$ = 1) closely resemble those of KV$_2$Se$_2$O [3], although Na$_{2-x}$V$_2$Se$_2$O crystallizes in an orthorhombic structure rather than a tetragonal lattice and the stacking layer (Na$^+$-layer) is quite different from the K$^+$-layer. These results suggest that the transport properties of $A$V$_2$Se$_2$O are to a large extent solely determined by the conducting [V$_2$Se$_2$O]$^{\delta-}$ layer. By replacing the type of the blocking layers, more novel materials with exotic magnetism or superconductivity could be realized in this structural family.

**SUMMARY & CONCLUSIONS**

Currently, the nature of magnetic interactions in Na$_{2-x}$V$_2$Se$_2$O remains unclear and will require further experimental and theoretical investigations to elucidate. Nevertheless, Na$_{2-x}$V$_2$Se$_2$O and KV$_2$Se$_2$O share very similar characteristics. Both contain the same structural unit, [V$_2$Se$_2$O]$^{\delta-}$, and exhibit magnetic phase transition temperatures above room temperature, along with SDW transitions at low temperatures. Accordingly, the emergence of superconductivity in Na$_{2-x}$V$_2$Se$_2$O is significant for a number of reasons. Firstly, as discussed before, this layered V-based oxychalcogenide combines structural features of both cuprate and iron-based superconductors. Furthermore, the 3$d^2$/3$d^3$ state in Na$_{2-x}$V$_2$Se$_2$O can be viewed as an electron-hole symmetric counterpart to the 3$d^8$/3$d^7$ state appearing in bilayer nickelate La$_3$Ni$_2$O$_7$. By comparing the properties of oxychalcogenide superconductors with those of the high-$T_C$ cuprates and nickelates, one can gain insight into the mechanism of superconductivity in layered oxides.

Secondly, vanadium exhibits a wide range of stable oxidation states, from 2$^+$ to 5$^+$, and many materials show mixed-valent behaviors accompanied by peculiar physical properties. For instance, the first superconductivity in vanadium oxides was found in the quasi-one-dimensional Na$_{0.33}$V$_2$O$_5$ under pressure, located in proximity to the charge-ordered antiferromagnetic insulator phase in the pressure-temperature phase diagram [61]. Recently, the discovery of superconductivity in $A$V$_3$Sb$_5$ ($A$=K, Rb, Cs) with Kagome structure has renewed interest in V-based compounds as potential unconventional superconductors [62]. In these materials, vanadium exists in a mixed valence state comprising V$^{4+}$ (3$d^1$) and V$^{5+}$ (3$d^0$). However, little is known about mixed valence vanadium oxides that involve the rare 2$^+$ oxidation state, except for the newly identified AMs, $A$V$_2$$Ch$$_2$O ($A$ = K, Rb, Cs; $Ch$ = Se, Te) [3-6]; and the geometrically frustrated spinel, AlV$_2$O$_4$ [63]. Sr$_2$VO$_4$ (V$^{4+}$, 3$d^1$), as one of the most promising candidates for high-$T_C$ superconductivity, has been extensively explored, yet no superconductivity was reported for this system until now. The discovery of superconductivity at a rather high critical temperature for the first time in such a low valence state makes V-based compounds of great significance as a platform for studying unconventional high-$T_C$ superconductivity.

Thirdly, while altermagnetism has been experimentally observed in numerous materials, critical questions remain regarding the types of superconductivity that might emerge in AM materials and whether they are topological. The parent Na$_{2-x}$V$_2$Se$_2$O stands out as a promising altermagnetic candidate due to its lack of inversion symmetry in the crystal lattice, distinguishing from the centrosymmetric altermagnet of KV$_2$Se$_2$O [3]. In general, a lack of spatial inversion symmetry in crystals can lead to a rich variety of physical phenomena. The nature of non-centrosymmetric altermagnets has not received much attention at present. In an altermagnet, when inversion symmetry is preserved, spin splitting exhibits an even-parity wave character (e.g., *d*-, or *g*-wave), as seen in the d-wave spin momentum locking in KV$_2$Se$_2$O. In contrast, a non-centrosymmetric system behaves an odd-parity state (e.g., *p*- or *f*-wave) due to the emergence of SOC [24,64,65]. While in a superconductor without inversion symmetry, parity-mixing can occur due to parity-breaking, enabling superconducting states to incorporate spin-triplet components [66-68]. Therefore, in Na$_{2-x}$V$_2$Se$_2$O, the unique combination of non-centrosymmetric structure and altermagnetism is expected to give rise to a distinctive superconducting pairing sate with an intriguing nodal structure, warranting further investigation. Our study opens new avenues for exploring unconventional superconductivity, particularly in systems that may involve possible topological states associated with spin-triplet pairing.

Finally, Na$_{2-x}$V$_2$Se$_2$O features a complex interplay of structural distortion, non-collinear SDW, AM，band topology, electronic correlation, and superconductivity, as revealed by our experimental results and preliminary DFT calculations. Since both Na$_{2-x}$V$_2$Se$_2$O and KV$_2$Se$_2$O host mixed-valent octahedral [V$_2$Se$_2$O]$^{\delta-}$ units, it is natural to consider the role of the blocking layer or interlayer coupling to explain the difference in their superconducting properties. Modifying the blocking layers allows precise tuning of crystal structure, electronic band topology, and other intriguing properties. Indeed, signs of superconductivity have been observed in some systems with different spacer layers [unpublished]. Therefore, further theoretical and experimental studies are needed to fully unleash the potential of materials containing [V$_2$Se$_2$O]$^{\delta-}$ structural unit, which provide larger space and challenge for exploring altermagnetic superconducting materials with higher critical temperature. The search for such systems is not only fundamental to advancing our knowledge of quantum materials but also holds potential for applications in low-dissipation spintronics and topological quantum computing.

# EXPERIMENTAL SECTION

**Synthesis of Na$_{2-x}$V$_2$Se$_2$O Polycrystalline Samples.** Polycrystalline samples of Na$_{2-x}$V$_2$Se$_2$O were synthesized via a conventional solid-state reaction method. The starting materials Na$_2$O, Na$_2$O$_2$, V, VSe and NaSe, were thoroughly mixed, pelletized, and sealed in tantalum tubes under an argon atmosphere. VSe was presynthesized by reacting V and Se at 650 °C for 24 h, while NaSe was synthesized from the elemental Na and Se in liquid ammonia. The sealed Ta tubes were jacketed in evacuated quartz ampoules to prevent oxidation and heated to 900 – 950 °C over 20 h, held at the target temperature for 20 h, and then naturally cooled to room temperature. To prevent the formation of superconducting impurity phases such as V$_3$Si caused by silicon introduction from quartz tubes, all sintering processes were strictly confined to tantalum containers or other metal tubes. To ensure phase homogeneity, the products were reground and sintered at 900 – 950 °C for an additional 24 h.

**Single Crystal Growth.** Single crystals of Na$_{2-x}$V$_2$Se$_2$O were grown using NaSe as the self-flux agent. Ground reactants were loaded into alumina crucibles, sealed in Ta tubes under argon, and jacketed within evacuated quartz ampoules. The assembly was heated to 950 °C over 20 h, held for 20 h, and then slowly cooled to 650 °C at a rate of 4°C /h before the furnace was turned off. The obtained samples are silvery and sensitive to air and moisture. Consequently, all sample handling was performed in an Ar-filled glove box with H$_2$O and O$_2$ levels maintained below 0.1 ppm.

**Topochemical Intercalation Synthesis of Na$_y$V$_2$Se$_2$O.** Na$_y$V$_2$Se$_2$O was also prepared by topochemical intercalation method. First, the parent compound V$_2$Se$_2$O was obtained through a deintercalation procedure. Polycrystalline KV$_2$Se$_2$O were synthesized from K, V$_2$O$_5$, V, and VSe$_2$ (pre-synthesized at 700 °C) via solid-state reaction. The resulting KV$_2$Se$_2$O crystals were immersed in excess deionized water at room temperature for one week to extract potassium, yielding black V$_2$Se$_2$O powders after drying under argon.

For the intercalation process, 100 mg of V$_2$Se$_2$O powder wasimmersed in 10 ml of tetrahydrofuran (THF) containing 1 mmol naphthalene and 1 mmol of sodium. The mixture was stirred at room temperature under an argon atmosphere. During the reaction, aliquots were extracted via pipetting, and the solid products were recovered by filtration.

**Structure and Composition Characterization.** Single-crystal X-ray diffraction data were collected on a BRUKER D8 VENTURE diffractometer. A specimen with approximate dimensions 0.2mm × 0.06 mm × 0.02mm was isolated and protected under an Ar atmosphere during data collection. The structure was solved and refined using the Bruker SHELXTL Software Package (more details in Supporting Information).

Powder X-ray diffraction (XRD) data were collected on a PANalytical diffractometer using Cu Kα radiation under ambient conditions. Air-sensitive samples were covered with Kapton tape inside a glove box prior to measurement. Crystal structure refinements were performed using the General Structure Analysis System (GSAS).

Elemental analysis was conducted using an energy-dispersive X-ray (EDX) spectrometer equipped with a Phenom scanning electron microscope (SEM) operated at 15 kV. Analysis was performed on a fresh surface cleaved from as-grown crystal, confirming the presence of Na, V, Se and O. Inductively coupled plasma (ICP) analysis was employed for a more precise determination of the sample composition. Both EDX and ICP analyses yielded consistent results, indicating a Na:V:Se ratio of approximately 1:2:2.

**Physical Property Measurements.** Electrical resistivity and Hall coefficient measurements were performed in a Quantum Design physical property measurement system (PPMS). Standard four-probe and Hall-bar method were adopted to obtain the resistivity and Hall coefficient, respectively. Magnetization measurements were carried out using a Quantum Design magnetic property measurement system (MPMS). Susceptibility-temperature curves were collected under both ZFC (zero field-cooled) and FC (field cooled) modes. To prevent deterioration in air, all samples used for physical property measurements were coated with N-grease.

**First-Principles Calculations.** The density functional theory (DFT) calculations were carried out using the Quantum ESPRESSO package [69]. The projector augmented wave pseudopotentials, based on generalized gradient approximation (GGA) exchange-correlation functionals of PBEsol, were applied [70,71]. Experimental crystallographic data were applied without relaxation. No magnetic order was considered. The Na vacancies were treated by removing one of the two Na atoms in the primitive cell (namely, $x = 1$ in $Na_{2-x}V_2Se_2O$). The energy cutoff for the wave functions was set to 60 Ry. Monkhorst-Pack $k$-point grids of $9 \times 9 \times 9$ and $13 \times 13 \times 11$ were used in the self-consistent, and density of states (DOS) calculations, respectively. The Fermi surfaces were constructed on a $101^3$ $k$-ponit grid, using the Wannier90 code [72].

**Acknowledgement**

This work was financially supported by the National Key R&D Program of China (Grant Nos. 2022YFA1403903 and 2025YFA1411501), the National Natural Science Foundation of China (Grant No. 12274440), the CAS Superconducting Research Project (No. SCZX-01011), and the Synergetic Extreme Condition User Facility (SECUF).

**Author Contributions**

Y. Sun, Z. Yin, and T. Zhang contributed equally to this work. Y. Sun, Z. Yin, T. Zhang, L. Wang, Y. Huang, J. He, W. Zhu, M. Ma, J. Bai, J. Cheng, Q. Dong, C. Li, P. Liu, Q. Liu, C. Zhang and G. Chen synthesized and characterized the samples, and carried out the magnetic and transport experiments. B. Ruan carried out the DFT calculations. All authors discussed the results. G. Chen conceived and supervised the project, analyzed the results and wrote the manuscript with inputs from all other authors.

**Competing interests**

The authors declare that they have no competing interests.

**Figures**

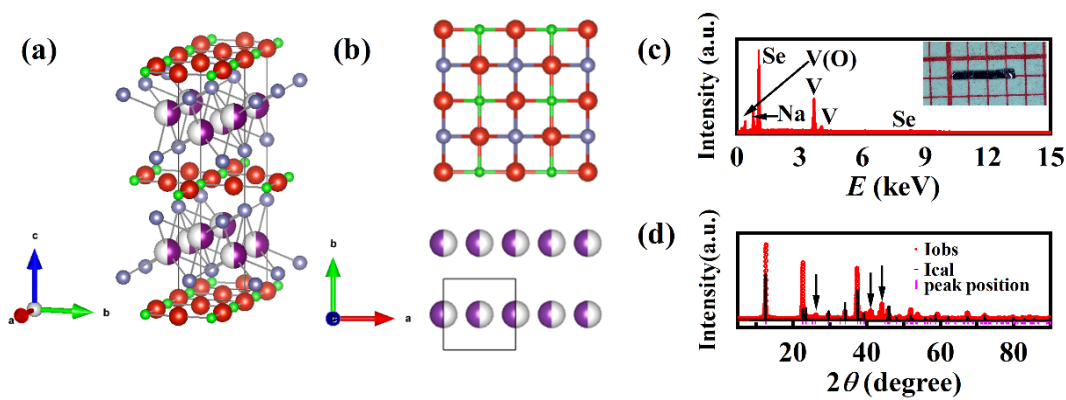

Figure 1: Crystal structure analysis of $Na_{2-x}V_2Se_2O$. (a) The crystal structure of $Na_{2-x}V_2Se_2O$ at room temperature. Purple, Green, red and blue spheres represent Na, O, V and Se atoms, respectively. (b) The structure of $V_2Se_2O$ layer (top) and the Na layer (bottom) viewed from the direction of $a$-axis. (c) The EDX spectrum of $Na_{2-x}V_2Se_2O$ single crystal. The average Na: V: Se atomic ratio is very close to 1: 2: 2, and no foreign elements were detected within the limitation of instrument resolution. (d) Powder XRD pattern of polycrystalline $Na_{2-x}V_2Se_2O$ sample at room temperature and corresponding refinement, except few peaks of impurity $Na_2Se$, all the peaks can be well indexed in a *Ammm* space group.

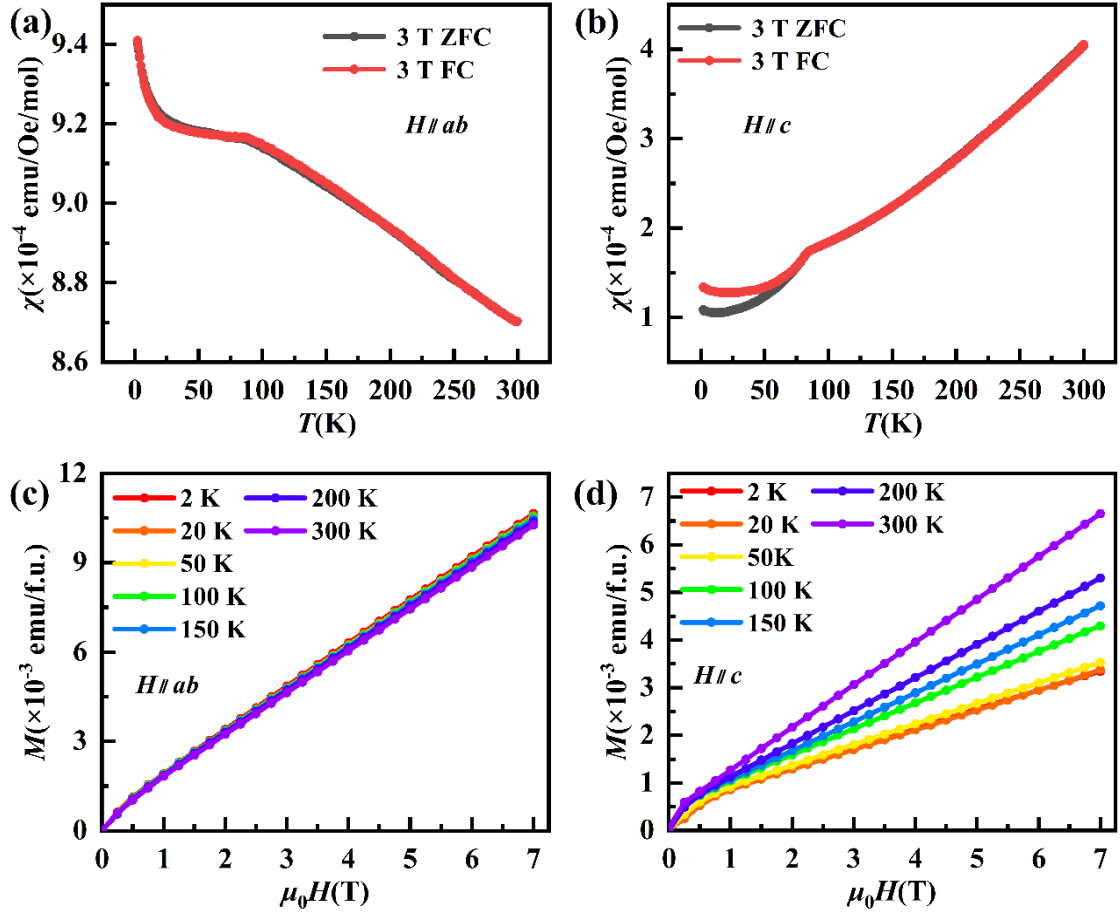

Figure 2: Magnetic characterization of non-superconducting parent $Na_{2-x}V_2Se_2O$ single crystals. (a, b) Temperature dependent magnetic susceptibility measured in zero-field-cooling (ZFC) and field-cooling (FC) modes with the magnetic field $\mu_0H$ = 3 T applied parallel to $c$-axis ($\mu_0H//c$) and $ab$-plane ($\mu_0H\perp c$), respectively. (c, d) The isothermal $M(\mu_0H)$s for both $\mu_0H//c$ and $\mu_0H\perp c$, respectively, measured at selected temperatures.

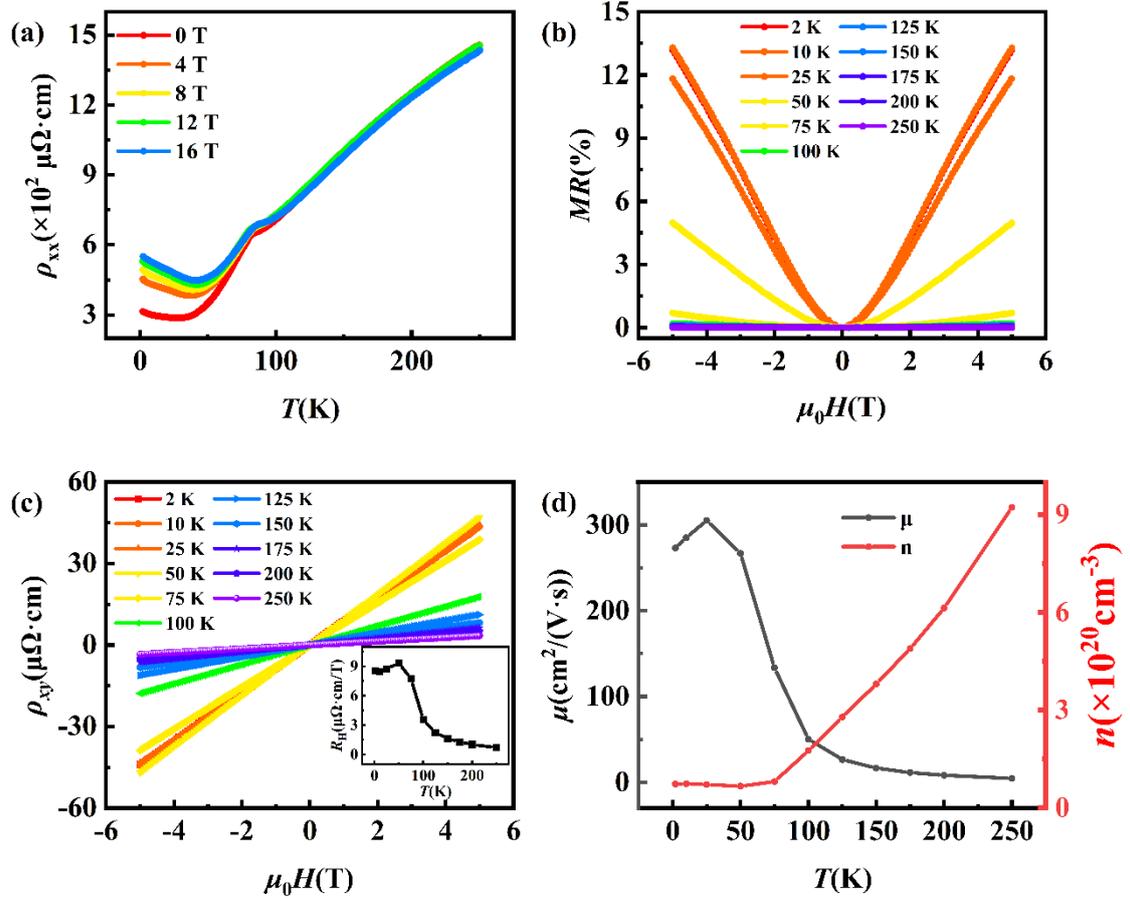

Figure 3: Transport properties of non-superconducting parent $Na_{2-x}V_2Se_2O$ single crystals. (a) Temperature dependence of the in-plane resistivity $\rho_{ab}(T)$ for a single crystal, measured at several magnetic fields. (b) Magnetoresistance (MR) as a function of magnetic field at selected temperatures. (c) Magnetic field dependence of the Hall resistivity at selected temperatures. The inset shows the temperature dependence of the Hall coefficient $R_H$. (d) Temperature dependence of carrier density n, and mobility μ extracted by a semiclassical one-band model.

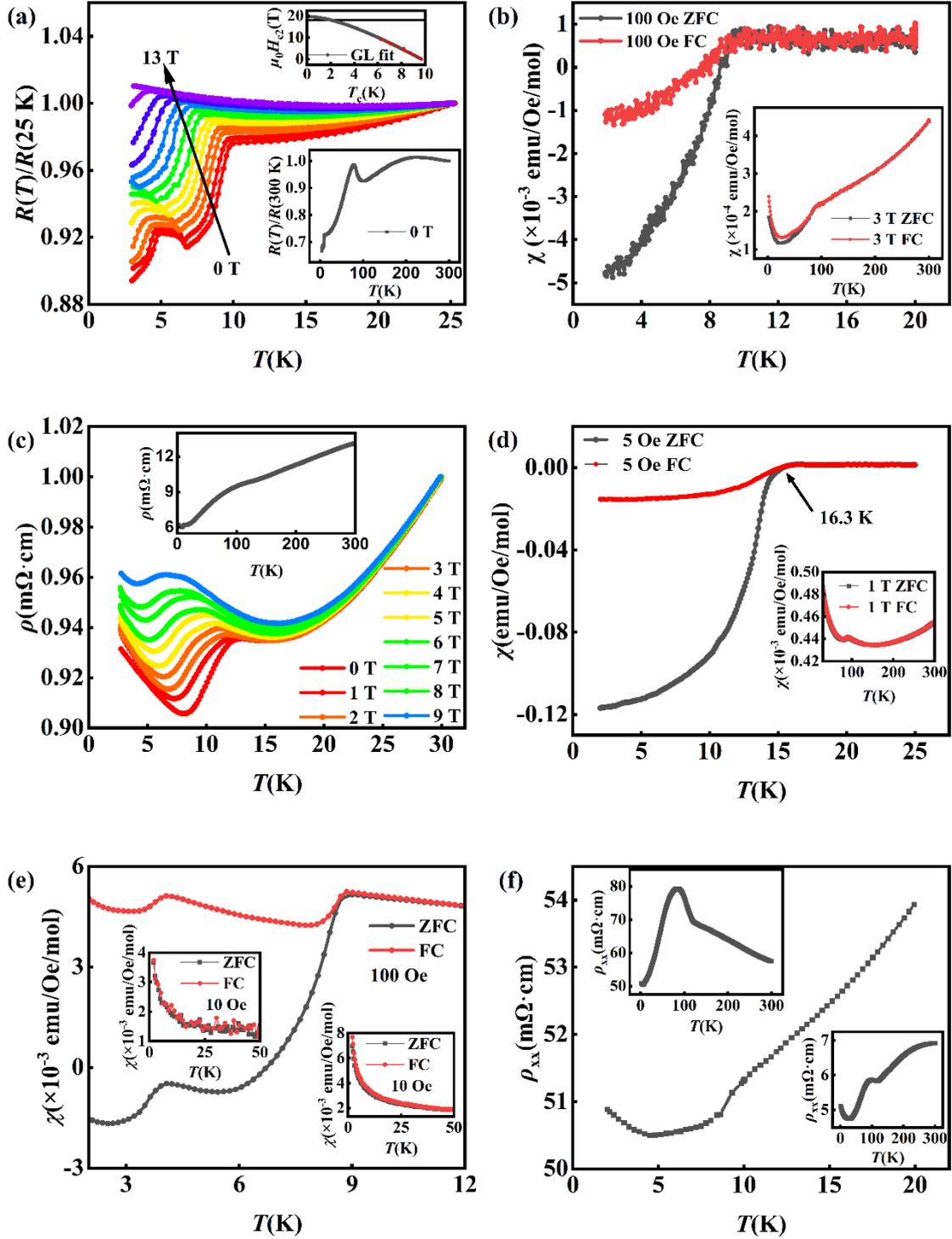

Figure 4: Superconducting properties of $Na_{2-x}V_2Se_2O$. (a) Resistivity vs temperature ($\rho-T$) behavior of a $Na_{2-x}V_2Se_2O$ single crystal under applied fields from 0 to 13 T in the superconducting region. Upper inset: Upper critical field fitted with Ginzburg-Landau (GL) equation. Lower inset: Zero-field $\rho(T)$ over 2 - 300 K. (b) Temperature dependence of the magnetic susceptibility $\chi(T)$ for a $Na_{2-x}V_2Se_2O$ single crystal under

ZFC and FC modes at 100 Oe in the superconducting region. Inset: $\chi(T)$ at 3 T (2 - 300 K). (c) $\rho(T)$ for a Na$_{2-x}$V$_2$Se$_2$O polycrystalline sample under fields of 0 - 9 T in the low temperature range. Inset: zero-field $\rho(T)$ over 2 - 300 K. (d) $\chi(T)$ for a Na$_{2-x}$V$_2$Se$_2$O polycrystalline sample under ZFC and FC modes at 5 Oe in the low temperature range. Inset: $\chi(T)$ at 1 T (30 - 300 K). (e) ZFC and FC $\chi(T)$ for Na-intercalated V$_2$Se$_2$O (18 h) at 10 Oe. Upper inset: $\chi(T)$ of pristine V$_2$Se$_2$O at 10 Oe. Lower inset: $\chi(T)$ of Na-intercalated V$_2$Se$_2$O for 36 hours at 10 Oe. (f) $\rho(T)$ of a single crystal after acetone immersion, showing a slight drop near 10 K. Upper-left inset: Zero-field $\rho(T)$ measured after soaking, spanning 2 - 300 K; lower-right inset: $\rho(T)$ measured before soaking.

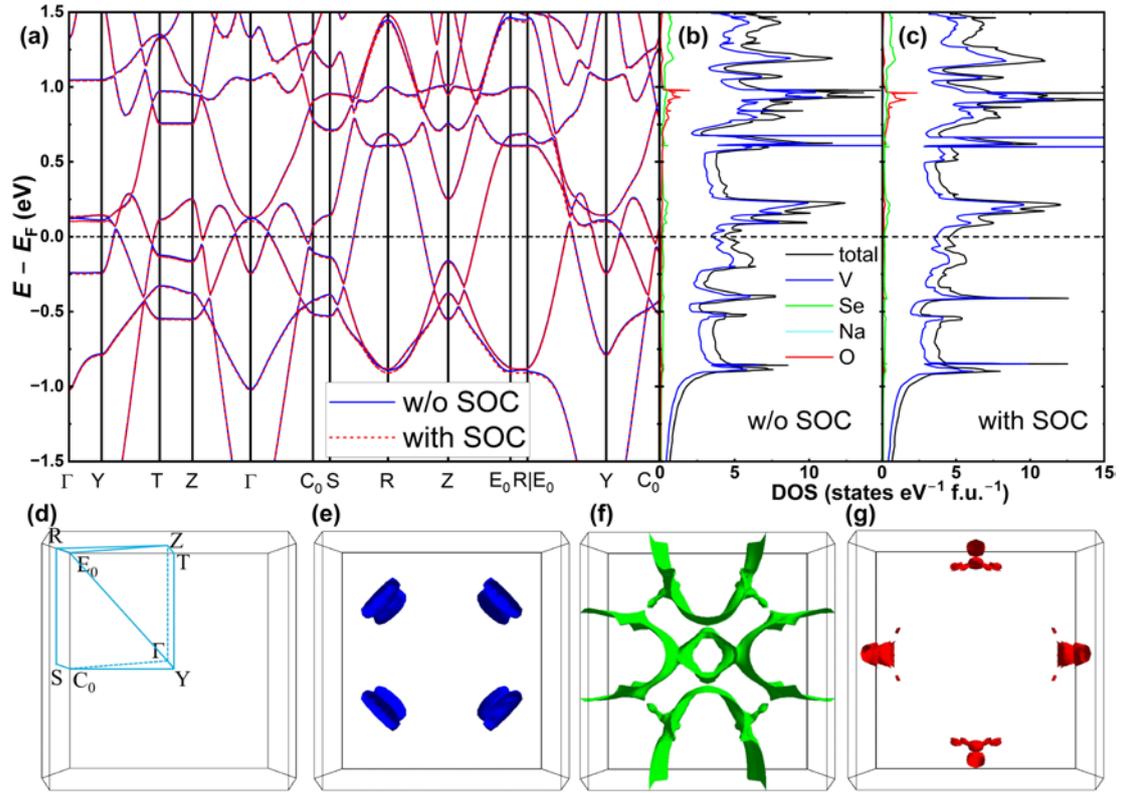

Figure 5: Calculated electronic structures of $Na_{2-x}V_2Se_2O$ ($x = 1$). (a) Electronic band structures with and without spin-orbit coupling (SOC). (b, c) Density of states (DOS) near the Fermi level with and without SOC, respectively. (d) The Brillouin zone with high symmetry points labeled. (e–g) DFT calculated Fermi surfaces.